\documentclass[%
 aip, apl,
 amsmath, amssymb,
 reprint,%
]{revtex4-2}

\usepackage{graphicx}
\usepackage{dcolumn}
\usepackage{bm}
\usepackage{xfrac}

\usepackage[utf8]{inputenc}
\usepackage[T1]{fontenc}
\usepackage{mathptmx}
\usepackage{etoolbox}

\makeatletter
\def\@email#1#2{%
 \endgroup
 \patchcmd{\titleblock@produce}
  {\frontmatter@RRAPformat}
  {\frontmatter@RRAPformat{\produce@RRAP{*#1\href{mailto:#2}{#2}}}\frontmatter@RRAPformat}
  {}{}
}%
\makeatother
\begin{document}

\title[draft]{Landauer-Limited Dissipation in Quantum-Flux-Parametron Logic}
\author{Quentin Herr}
 \email{quentin.herr@imec-int.com}
 \affiliation{imec USA, Kissimmee, FL 34744 USA}
\date{\today}

\begin{abstract}
The Landauer limit is to irreversible logic what the Carnot cycle is
to heat engines. This limit is approached in the adiabatic Quantum
Flux Parametron (aQFP) by copying the inputs of standard logic gates
to produce reversible logic gates, and disposing of the copied inputs
using the ``terminate'' gate dissipating only the thermal energy, $\ln
2\,kT$. This method eliminates the non-adiabatic switching associated
with backaction that arises in conventional aQFP logic. Real aQFP
devices have parameter mismatch causing proportionate increases in
dissipation and bit-error rate. A chip with $10^9$ aQFPs with
realistic fabrication spread of 1\%-$1\sigma$ control on junction
critical current and 5\%-$1\sigma$ on inductors would have outlier
devices with a bit-error rate of $10^{-31}$, compared to $10^{-71}$
for ideal devices. Power dissipation across all devices on-chip would
increase to about 7$\times$ the Landauer limit. An ideal circuit
processing correlated bit streams dissipates fractional bit energy per
cycle commensurate with the information lost, in accord with
Landauer's concept of logical entropy.

\end{abstract}

\maketitle Irreversible logic operating in the Landauer limit
\cite{landauer1961irreversibility} would be an attractive design
point, offering energy efficiency for conventional logic
functions. Ideally, the rate of thermally-induced bit errors is
completely decoupled from energy dissipation. Bit-error rate
scales with the (arbitrary) height of the potential barrier in a
bistable potential well, and energy dissipation scales with the
thermal energy. The original literature observes that ``these
considerations do not assure us that the required potentials are
physically realizable.''\cite{keyes1970minimal}

The Parametric Quantron (PQ) \cite{likharev1977dynamics1,
  likharev1982classical} and adiabatic Quantum Flux Parametron
(aQFP)\cite{takeuchi2012margin} are Josephson superconducting logic
gates that adhere closely to the ideal system. The exception is
backaction during reset, when the state is in the upper potential well
and escapes into the lower well dissipating energy several orders
above the Landauer limit. These gates do not support ``the thin
barrier at the center of the well'' specified for the ideal system
\cite{keyes1970minimal}. A reversible aQFP logic function has been
described that avoids backaction \cite{takeuchi2017reversibility}, as
the state never finds itself in the upper well. Unwanted outputs of
the reversible aQFP gate can be thermalized using a ``terminate'' gate
\cite{yamae2024minimum} with power dissipation just that of Landauer's
limit.

This paper shows how standard logic gates can be made reversible by
copying the inputs (consistent with
\onlinecite{likharev1982classical}), and Landauer's limit can be
achieved by thermalizing the copies. Energy dissipation and bit-error
rates are modeled both for ideal devices and for devices with
parameter mismatch. Various circuit configurations serve to explore
the continuum between reversible and Landauer-limited
operations. Important operations such as serial and parallel fanout
are shown to be dissipation free. Logical processing of correlated bit
streams dissipates energy in proportion to information loss,
consistent with Landauer's original concept of logical entropy.

\section*{Context-Dependent Gate Operations}
The aQFP has a single thermodynamic degree of freedom, superconductor
phase. An isomorphic mechanical model consists of two
physical pendulums connected by a stiff torsion spring. Input and
output signals produce equal torque on the pendulums; the clock
produces equal-and-opposite torque. Superconductor phase corresponds
to the mean angular displacement of the pendulums.

\begin{figure*}
  \includegraphics[width=6.3in]{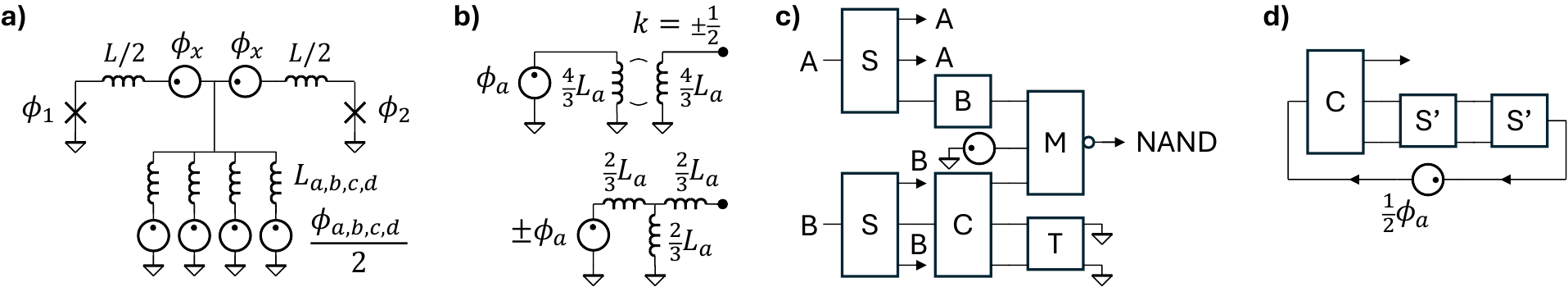}
  \caption{\label{fig1} aQFP circuit schematics. a) Four parallel
    inductors $L_{a\mbox{-}d}$ are connected to input and output aQFPs
    that are modeled as superconducting phase sources. The aQFP is
    clocked and powered by the exciter phase source $\phi_x$, which is
    implemented via transformer coupling to the clock. b) Transformer
    coupling is also used between stages, which attenuates phase
    amplitude by a factor of two. The sign of the coupling constant
    (top) determines the polarity of the coupling, with polarity
    inversion producing logical negation. An equivalent circuit model
    (bottom) for positive coupling shows the equivalence to the loop
    inductance and attenuation used in (a). c) A NAND gate consists of
    aQFP stages with each column driven by a different phase of the
    multi-phase clock. Both inputs are connected to splitters S to
    achieve fanout. The upper half uses a buffer B that ensures stable
    operation in the presence of backaction from the MAJ gate M. The
    lower half uses a copy gate C that fans out to MAJ and to a
    termination gate T. The buffer dissipates energy comparable to the
    signal due to non-adiabatic switching on reset, but terminate
    dissipates only thermal energy $\ln2\,kT$, consistent with
    Landauer's limit. d) The central, constant logical input to the
    MAJ gate is generated by a constant-phase source applied in
    series with a multi-stage loop that produces a signal like the
    other inputs. The polarity of the phase source determines whether
    the input is logical ``0'' or ``1.''}
\end{figure*}

\begin{figure*}
  \includegraphics[width=7.0in]{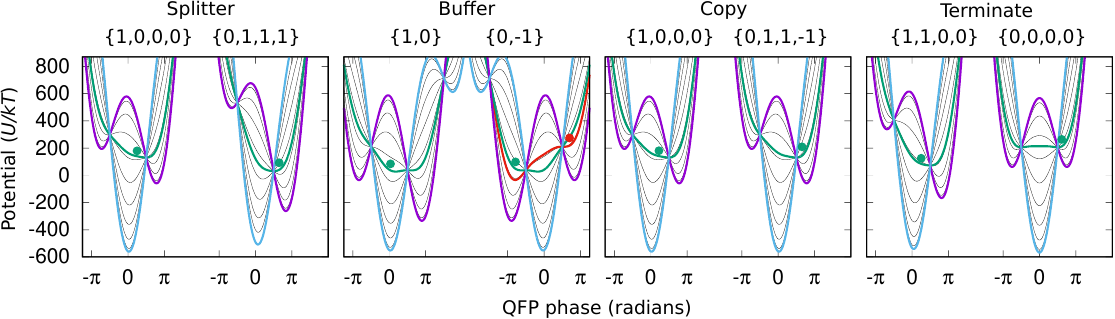}
\caption{\label{fig2} Evolution of the aQFP potential as a function of
  aQFP phase, $\phi_+$, when the exciter phase, $\phi_x$, is ramped
  from 0 to $\pi$ (left half of each frame) and ramped back down
  (right half). The blue line corresponds to exciter phase of 0 and
  the purple line to $\pi$. Thin lines are intermediate values equally
  spaced in $\phi_x$. Green and red lines are the intermediate values
  that have zero slope at the saddle points where all curves
  cross. Each frame is labeled with the digital values of the
  inputs/outputs and the corresponding logical functions. For
  illustration, particles that correspond to aQFP phase are shown
  slightly above the minimum energy and are vectored towards the
  minimum. For clarity, only switching to the right-side well is
  shown.}
\end{figure*}

Circuit schematics of the aQFP are shown in Fig.~\ref{fig1}.  Typical
aQFP circuits consist of splitter-buffer-majority stages
\cite{ayala2020mana}.  The same model applies to all use cases, which
differ only by the state-dependent couplings to all neighbors.
Whereas early modeling treated gate inputs as current sources
\cite{takeuchi2012margin, ko1992noise}, both input and output
couplings need to be included in a model of interconnected logic
gates.

Switching events for the aQFP in various contexts are shown in
Fig.~\ref{fig2}. The equations used to generate the potentials are
given in the supplemental material. The model uses established
parameter values for the Josephson junction critical current and
interconnect inductances \cite{takeuchi2012margin},
$I_c=50\,\mu\rm{A}$, $L=1.3\,{\rm pH}$, and $L_{a\mbox{-}d}=42\,{\rm
  pH}$, and the potential is normalized to the thermal energy at
$T=4.2\,K$. The excited aQFP produces a superconductor phase amplitude
of 2.56\,rad, corresponding to $\pm20\,\mu{\rm A}$ supplied to each
output. Signal inputs are attenuated in the interconnect to $\pm10\,\mu{\rm A}$.

The buffer stage is required for stability in case the signal finds
itself in the minority at the input to the MAJ gate. Input signals in
the majority determine the switching polarity of the MAJ gate, but a
signal in the minority sees backaction from the MAJ gate that {\it
  doubles} the signal amplitude, placing the buffer in the higher of
the two potential wells, leading to irreversible, non-adiabatic
switching. A way to see the irreversibly of this event is to run time
backwards after the switching event has occurred. The buffer would not
end up where it started but would instead find the lower potential
well.

\begin{figure*}[tbh]
\includegraphics[width=7.0in]{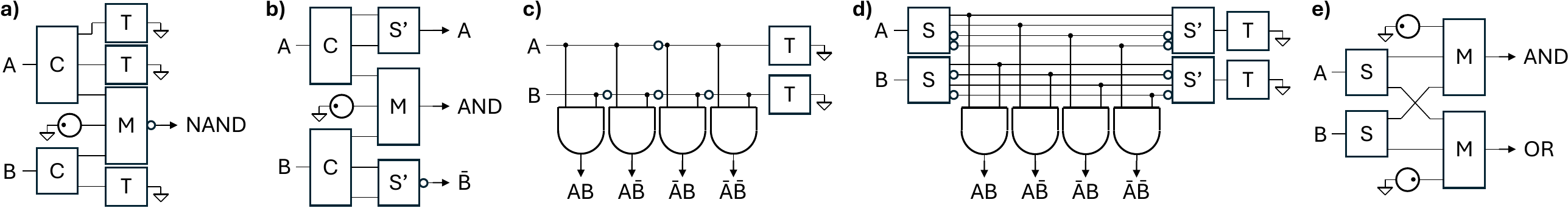}
\caption{\label{fig3} Various circuit configurations that eliminate
  non-adiabatic switching using the terminate function. a) The upper
  half of the NAND gate uses two independent terminations, which
  doubles the dissipated energy. The lower half uses a pruned copy
  gate with a single connection to terminate, but this copy gate itself
  will dissipate when it is the minority input to MAJ. b) The copied
  outputs can be wired out instead of terminated, and selectively
  inverted. c) A two-to-four address decoder is implemented by daisy
  chaining the gates of (b), with only two terminations for the whole
  circuit. d) The same decoder function is implemented by broadcasting
  the address inputs in parallel to all gates. The parallel signals
  can be recombined, again using only two terminations for the whole
  circuit. e) A circuit that generates logical AND and OR uses a pair
  of splitters wired directly to both MAJ gates. The splitters
  dissipate the thermal energy only when the input signals differ.}
\end{figure*}

The lower half of the NAND gate in Fig.~\ref{fig1} has a splitter-{\it
  copy}-majority configuration. When the copy gate is the minority
input to the MAJ gate the backaction will be outweighed by the two
connections to the termination. The terminate gate has its outputs
shorted to ground, so all inputs and outputs are zero after the input
is deasserted and the gate exhibits a symmetric double-well
potential. As the barrier between wells is lowered, the state will
escape from a single well and spend time occupying both wells before
being thermalized by reconfinement to a the single well. The
thermodynamic cycle for thermalization is sudden expansion followed
by isothermal compression. Sudden expansion increases the entropy by
$\ln 2\,k$, like doubling the volume of a gas
\cite{keyes1970minimal}, and isothermal compression does work on the
particle that is rejected as heat, in exchange for decreased logical
entropy.

The aQFPs in this system never occupy the upper well, but they all
occupy a symmetric double during that part of the cycle when the
inputs have been reset and the aQFPs connected to the outputs have not
yet been excited. The symmetric double wells are as shown for the
terminate in Fig.~\ref{fig2}. The aQFP must hold non-equilibrium state
during this part of the cycle, with bit-error rate (BER) induced by
thermal activation over the barrier given by the Kramers escape
rate. Barrier height in the current design is about $\Delta U/kT=500$,
which is functionally equivalent to the arbitrarily-high barrier
specified in Ref.~\onlinecite{keyes1970minimal}.

\section*{Physically Realizable Nonideal Devices}
The aQFP is a real device, with practical considerations including
parameter mismatch and the energy efficiency in a closed system
including the clock source. Parameter mismatch increases the BER of
conventional aQFP logic operations and increases the dissipation of
the terminate. A chip with $10^9$ aQFPs fabricated in a process with
$1\sigma$ control of 1\% on junction critical current and $1\sigma$
control of 5\% on inductors is considered in the supplemental
material. Dissipation for each terminate would increase by about
7$\times$ amounting to ``15\% Landauer efficiency''. The potential
barrier $\Delta U$ plotted in Fig.~\ref{fig2} would be reduced by a
factor of 0.43 for the outliers, but would still produce a chip-wide
error rate better than $10^{-23}$. In this way non-ideal aQFPs
experience graceful degradation of performance. By contrast, in the
billiard ball model of reversible computing
\cite{fredkin1982conservative} nonideality results in outright
failure.

Much work done is by the clock when the particle is lifted, but is
returned to the source when the particle is lowered again, with
exceptions for non-adiabatic switching in the buffer and isothermal
compression in the terminate. These events cannot be thermalized by a
clock generated at e.g. room temperature. Instead, the aQFPs need to
be well-isolated thermally from the higher temperature, which is
accomplished by weak coupling and narrow clock bandwidth. The heat
generated by the aQFP must be scaled by the cryocooler efficiency to
find wall-plug dissipation, but a clock generator rejecting heat at
room temperature makes only a small contribution to the total power
budget. This all-electrical system is quite efficient compared to the
opto-fluidic system reported in
Ref.~\onlinecite{berut2012experimental}, where dissipation in the
control is many orders above the thermal energy.

\section*{Power Dissipation and Logical Entropy}
The various circuit configurations shown in Fig.~\ref{fig3} inform the
relationship between power dissipation and logical entropy. Two
terminations in parallel are the functional equivalent of a single
terminate. However, use of multiple devices increases the
thermodynamic degrees of freedom and dissipation. A copy gate
connected to a MAJ input as in Fig.~\ref{fig1} but with a {\it single}
connection to terminate avoids non-adiabatic switching due to
backaction, but dissipates the thermal energy in the copy gate in case
of backaction.

The copied inputs need not be terminated if they can instead be reused
by subsequent logic gates. Daisy-chaining the outputs from one stage
to the inputs of the next as in a two-to-four address decoder produces
a schematic like those of reversible circuits. In this configuration
the inputs are used multiple times, but need to be terminated only
once. The inputs can instead fan out to the gates in parallel, and the
entropy of the parallel signals, including negations, is still that of
a single bit. The signals can be recombined, which involves a splitter
operating in reverse, and sent to a single terminate. This is possible
because copying the bits and deleting the copies are logically
reversible operations\cite{bennett2003notes}.  In the reversible
circuit reported in Ref.~\onlinecite{takeuchi2017reversibility}, it is
auxiliary gate outputs rather than copies of gate inputs that are
terminated.

The circuit shown in Fig.~\ref{fig3}e produces logical AND and
OR. This gate dissipates the thermal energy in the splitters on those
cycles when the inputs differ. On the physical level, differing inputs
cause differing outputs from the MAJ gates. The digital values of
inputs and outputs of the splitters during reset are then \{0,1,-1\} and
\{0,-1,1\}. These signals sum to zero, and the reset cycle is like that
of a termination. When the inputs differ one bit of information is
lost; the inputs cannot be fully reconstructed from the outputs, but
are known to differ. The reset cycle in this case thermalizes one bit
in {\it each} of the two splitters. When the inputs are equal there is
no backaction, the gate does not dissipate, and no information is
lost.

\begin{table}[tbh]
\caption{\label{table1} OR-AND Logical Entropy and Dissipation}
\begin{ruledtabular}
\begin{tabular}{cccc|cccc|c}
A & B & $P_i$ & $S_i/k$ & AB & A+B & $P_f$ & $S_f/k$ & $Q/kT$ \\
\hline
0 & 0 & $\alpha\!=\!\sfrac{1}{2}\!-\!2\epsilon$ & $\alpha\ln\sfrac{1}{\alpha}$ & 0 & 0 & $\alpha$ & $\alpha\ln\sfrac{1}{\alpha}$ & 0 \\
0 & 1 & $\epsilon$ & $\epsilon\ln\sfrac{1}{\epsilon}$ & 0 & 1 & $2\epsilon$ & $2\epsilon\ln\sfrac{1}{2\epsilon}$ & $2\epsilon\ln2$ \\
1 & 0 & $\epsilon$ & $\epsilon\ln\sfrac{1}{\epsilon}$ & - & - & - & - & $2\epsilon\ln2$ \\
1 & 1 & $\sfrac{1}{2}$ & $\sfrac{1}{2}\ln2$ & 1 & 1 & $\sfrac{1}{2}$ & $\sfrac{1}{2}\ln2$ & 0 \\
\end{tabular}
\end{ruledtabular}
\end{table}

Consider a toy error correction scheme effective when the receiver
threshold is set too low and a logical ``0'' sometimes registers as a
logical ``1.'' The bit stream is duplicated and the OR-AND gate is
used to correct errors. Following Landauer's method
\cite{landauer1961irreversibility}, the entropy of this system is
given in Table.~\ref{table1} for error probability $\epsilon$. The
pairs of bits are perfectly correlated and contain a single bit of
information apart from bit errors. The reduction in logical entropy,
given as the difference between initial and final values, is $S_i-S_f
= 2\epsilon\ln\sfrac{1}{\epsilon}\,k-2\epsilon\ln\sfrac{1}{2\epsilon}\,k
= 2\epsilon\ln2\,k$. Dissipation in the OR-AND circuit is just twice
the Landauer limit, and scales to arbitrarily low values as the error
probability goes to zero.

The Landauer-limited aQFP is a physically-realizable device serving as
an effective vehicle to explore the ultimate limits of
computation. Modest clock rates, high latency, and high clock
amplitude arising from weak coupling limits practical applications in
high-performance digital logic. Computational density scaling
expressed as operations per second per unit chip area would be about
1000$\times$ less than that of conventional SFQ
circuits~\cite{herr2023superconducting}. The device may yet have
practical value at mK temperatures as auxiliary circuits for qubit
control and readout\cite{berkley2010scalable}.

\begin{acknowledgments}
The work was supported by Osceola County, FL.
\end{acknowledgments}

\section*{Supplemental Material}
The aQFP potential is given by \cite{ko1992noise, takeuchi2012margin}
\begin{equation}
  U=\frac{I_c\Phi_0}{2\pi}\left[\frac{(\phi_x-\phi_-)^2}{\beta_L}
    +\frac{(\phi_{\,\rm in}-\phi_+)^2}{\beta_L+2\beta_q}-2\cos\phi_-\cos\phi_+\right],
  \label{U}
\end{equation}
where $\phi_+=\sfrac{1}{2}(\phi_1+\phi_2)$ is the mean phase of the
Josephson junctions of the aQFP,
$\phi_-=\sfrac{1}{2}(\phi_1-\phi_2)$ is one-half the phase difference,
and $\phi_x$ is the exciter phase. The normalized inductances are
$\beta_L=2\pi LI_c/\Phi_0$ and $\beta_q=2\pi L_qI_c/\Phi_0$, where
$L_q$ is the parallel combination of the input and output interconnect
inductors.

To plot the potential of the aQFP in two dimensions, the value
$\phi_{-,{\rm min}}$ that minimizes the potential can be expressed in
terms of the other phase parameters. Setting $dU/d\phi_-=0$ yields
$\phi_x-\phi_-=\beta_L\sin\phi_-\cos\phi_+$. For $\beta_L \ll
2\beta_q$, $\phi_-$ is about equal to
$\phi_x$\cite{takeuchi2012margin}. Substituting
$\phi_-=\phi_x+\epsilon$ and to first-order $\sin(\phi_x+\epsilon)
\simeq \sin\phi_x+\epsilon\cos\phi_x$ yields
\begin{equation}
  \phi_{-,{\rm min}}\simeq
  \phi_x-\frac{\beta_L\sin\phi_x\cos\phi_+}{1+\beta_L\cos\phi_x\cos\phi_+}.
  \label{pm}
\end{equation}
In Fig.~\ref{fig2}, the function $U(\phi_+)$ is plotted for different
values of exciter $\phi_x$ and signal $\phi_{\,\rm in}$ after
substituting Eq.~\ref{pm} into Eq.~\ref{U}.

The curves in Fig.~\ref{fig2} cross each other at
$\phi_+=\pm\pi/2$. The curve with its minimum at the crossing can be
plotted by finding the corresponding value of $\phi_x$. Setting
$dU/d\phi_+=0$ yields
\[
  \phi_{x\rm,cross}=\arccos\left(\frac{\pm\phi_{\,\rm in}-\pi/2}{\beta_L+2\beta_q}\right).
  \]
  
In the adiabatic limit, the gate is in equilibrium throughout the
switching event and the BER is given by the probability of occupying
the upper well, ${\rm BER}=\sfrac{1}{2}{\rm erfc}\,(-\Delta U/kT)$. A simple
estimate for $\Delta U$, the energy difference between wells, is given
by the points at $\phi_+=\pm\pi/2$ where all the curves cross. The
potential difference between the points at the crossings,
\begin{equation}
  \Delta U_{\rm cross}=
    \alpha\frac{I_c\Phi_0\phi_{\,\rm in}}{\beta_L+2\beta_q}=
    \alpha\frac{\Phi_0^2}{L+2L_q}\frac{\phi_{\,\rm in}}{2\pi},
  \label{Ucross}
\end{equation}
where $\Phi_0=h/(2e) \simeq 2.07\,{\rm mA\,pH}$ is the
Single-Flux-Quantum (SFQ). The factor $\alpha$ is unity for
well-targeted parameter values. This estimate is equivalent to the
$\sfrac{1}{2}LI^2$ energy of the signals in the interconnect. (In an simple
approximation, a signal in one of the input inductors is completely
absorbed and the amplitude nulled under correct switching, but the
amplitude is doubled under incorrect switching.) The splitter and the
MAJ gate with mixed inputs both have a signal strength $\phi_{\,\rm
  in}=2.56/8$, which gives $\Delta U/kT=168$ and a negligible BER of
$10^{-71}$.

Elevated BER in the logic gates due to device parameter mismatch can
be quantified. If one input inductor where mistargeted 33\% low and the
other two 33\% high, a minority signal on the smaller inductor would
have equal weight to opposite-polarity signals on the other two
inductors, causing the circuit to fail outright. In similar fashion,
if the junction critical currents were mistargeted in opposite
directions by $5\,\mu{\rm A}$ each, the threshold would shift by
$10\,\mu{\rm A}$, equal to the input signal and again causing outright
failure. Smaller excursions will produce a linear decrease in $\Delta
U$, so the prefactor in Eq.~\ref{Ucross} is
\begin{equation}
  \alpha=5(\delta i_c-\delta i_c\,\!\!\!')
    +\frac{a}{1+\delta l_a}+\frac{b}{1+\delta l_b}+\frac{c}{1+\delta l_c}+\frac{d}{1+\delta l_d},
  \label{amis}
\end{equation}
where $\delta$ denotes the parameter excursion normalized to the
nominal value. The set of numbers \{a,b,c,d\} have values $\pm1$, or 0,
denoting the digital values of the aQFP inputs and outputs.

Consider a chip with $10^9$ aQFPs fabricated in a process with $1\sigma$
of 1\% control on junction critical current $1\sigma$ of 5\% on
inductors. Several devices would be expected to have $6\sigma$
excursions. However, the $6\sigma$ event with the worst-case impact
has additive $2.3\sigma$ excursions on all five circuit components. As
there are two ways to choose junction mismatch and then three ways to
choose worst case inductor mismatch,
\[
  \sfrac{1}{2}\,\rm{erfc}\,(\sfrac{6}{\sqrt{2}}) \simeq
  2\cdot3\cdot\left[\sfrac{1}{2}\,\rm{erfc}\,(\sfrac{2.3}{\sqrt{2}})\right]^5 \simeq
  10^{-9}.
  \]
This worst-case parameter set will lower the $\Delta U$ of
Eq.~\ref{Ucross} by a factor given in Eq.~\ref{amis} of $\alpha =
5(-0.023 \cdot 2)\!+\!2/1.115\!-\!1/0.885\!=\!0.43$. $\Delta U$ is
reduced to less than half the well-targeted value, but still indicates
a very low $\rm{BER}=\exp(-73) \simeq 10^{-32}$ per device and better
than $10^{-23}$ per chip.

Parameter mismatch of the Josephson junctions tilts the double well
potential slightly, causing non-adiabatic switching and elevated
energy dissipation in the terminate. Inductor mismatch is unimportant
as all signals are zero, and it is the mean value of Josephson
junction device mismatch that matters here. For normally distributed
parameters with standard deviations of $\sigma$, the expectation value
\[
  \lvert \delta i_c-\delta i_c\,\!\!\!' \rvert \simeq 1.128\sigma/I_c.
\]
Referring again to Eq.~\ref{Ucross} and Eq.~\ref{amis}, $\alpha = 0.0565$ and 
$\Delta U_{\rm mis} = 9\,kT$. The terminate will dissipate this
energy only half the time, when the particle finds itself in the upper
well. Overall, the terminate could be said to dissipate a mean value
of $4.5\,kT$, and the circuit to operate at 15\% of Landauer efficiency.

\bibliography{pq}

\begin{thebibliography}{14}%
\makeatletter
\providecommand \@ifxundefined [1]{%
 \@ifx{#1\undefined}
}%
\providecommand \@ifnum [1]{%
 \ifnum #1\expandafter \@firstoftwo
 \else \expandafter \@secondoftwo
 \fi
}%
\providecommand \@ifx [1]{%
 \ifx #1\expandafter \@firstoftwo
 \else \expandafter \@secondoftwo
 \fi
}%
\providecommand \natexlab [1]{#1}%
\providecommand \enquote  [1]{``#1''}%
\providecommand \bibnamefont  [1]{#1}%
\providecommand \bibfnamefont [1]{#1}%
\providecommand \citenamefont [1]{#1}%
\providecommand \href@noop [0]{\@secondoftwo}%
\providecommand \href [0]{\begingroup \@sanitize@url \@href}%
\providecommand \@href[1]{\@@startlink{#1}\@@href}%
\providecommand \@@href[1]{\endgroup#1\@@endlink}%
\providecommand \@sanitize@url [0]{\catcode `\\12\catcode `\$12\catcode
  `\&12\catcode `\#12\catcode `\^12\catcode `\_12\catcode `\%12\relax}%
\providecommand \@@startlink[1]{}%
\providecommand \@@endlink[0]{}%
\providecommand \url  [0]{\begingroup\@sanitize@url \@url }%
\providecommand \@url [1]{\endgroup\@href {#1}{\urlprefix }}%
\providecommand \urlprefix  [0]{URL }%
\providecommand \Eprint [0]{\href }%
\providecommand \doibase [0]{https://doi.org/}%
\providecommand \selectlanguage [0]{\@gobble}%
\providecommand \bibinfo  [0]{\@secondoftwo}%
\providecommand \bibfield  [0]{\@secondoftwo}%
\providecommand \translation [1]{[#1]}%
\providecommand \BibitemOpen [0]{}%
\providecommand \bibitemStop [0]{}%
\providecommand \bibitemNoStop [0]{.\EOS\space}%
\providecommand \EOS [0]{\spacefactor3000\relax}%
\providecommand \BibitemShut  [1]{\csname bibitem#1\endcsname}%
\let\auto@bib@innerbib\@empty
\bibitem [{\citenamefont {Landauer}(1961)}]{landauer1961irreversibility}%
  \BibitemOpen
  \bibfield  {author} {\bibinfo {author} {\bibfnamefont {R.}~\bibnamefont
  {Landauer}},\ }\bibfield  {title} {\enquote {\bibinfo {title}
  {Irreversibility and heat generation in the computing process},}\ }\href@noop
  {} {\bibfield  {journal} {\bibinfo  {journal} {IBM journal of research and
  development}\ }\textbf {\bibinfo {volume} {5}},\ \bibinfo {pages} {183--191}
  (\bibinfo {year} {1961})}\BibitemShut {NoStop}%
\bibitem [{\citenamefont {Keyes}\ and\ \citenamefont
  {Landauer}(1970)}]{keyes1970minimal}%
  \BibitemOpen
  \bibfield  {author} {\bibinfo {author} {\bibfnamefont {R.~W.}\ \bibnamefont
  {Keyes}}\ and\ \bibinfo {author} {\bibfnamefont {R.}~\bibnamefont
  {Landauer}},\ }\bibfield  {title} {\enquote {\bibinfo {title} {Minimal energy
  dissipation in logic},}\ }\href@noop {} {\bibfield  {journal} {\bibinfo
  {journal} {IBM Journal of Research and Development}\ }\textbf {\bibinfo
  {volume} {14}},\ \bibinfo {pages} {152--157} (\bibinfo {year}
  {1970})}\BibitemShut {NoStop}%
\bibitem [{\citenamefont {Likharev}(1977)}]{likharev1977dynamics1}%
  \BibitemOpen
  \bibfield  {author} {\bibinfo {author} {\bibfnamefont {K.}~\bibnamefont
  {Likharev}},\ }\bibfield  {title} {\enquote {\bibinfo {title} {Dynamics of
  some single flux quantum devices: I. {P}arametric quantron},}\ }\href@noop {}
  {\bibfield  {journal} {\bibinfo  {journal} {IEEE Transactions on Magnetics}\
  }\textbf {\bibinfo {volume} {13}},\ \bibinfo {pages} {242--244} (\bibinfo
  {year} {1977})}\BibitemShut {NoStop}%
\bibitem [{\citenamefont {Likharev}(1982)}]{likharev1982classical}%
  \BibitemOpen
  \bibfield  {author} {\bibinfo {author} {\bibfnamefont {K.~K.}\ \bibnamefont
  {Likharev}},\ }\bibfield  {title} {\enquote {\bibinfo {title} {Classical and
  quantum limitations on energy consumption in computation},}\ }\href@noop {}
  {\bibfield  {journal} {\bibinfo  {journal} {International Journal of
  Theoretical Physics}\ }\textbf {\bibinfo {volume} {21}},\ \bibinfo {pages}
  {311--326} (\bibinfo {year} {1982})}\BibitemShut {NoStop}%
\bibitem [{\citenamefont {Takeuchi}\ \emph {et~al.}(2012)\citenamefont
  {Takeuchi}, \citenamefont {Ehara}, \citenamefont {Inoue}, \citenamefont
  {Yamanashi},\ and\ \citenamefont {Yoshikawa}}]{takeuchi2012margin}%
  \BibitemOpen
  \bibfield  {author} {\bibinfo {author} {\bibfnamefont {N.}~\bibnamefont
  {Takeuchi}}, \bibinfo {author} {\bibfnamefont {K.}~\bibnamefont {Ehara}},
  \bibinfo {author} {\bibfnamefont {K.}~\bibnamefont {Inoue}}, \bibinfo
  {author} {\bibfnamefont {Y.}~\bibnamefont {Yamanashi}},\ and\ \bibinfo
  {author} {\bibfnamefont {N.}~\bibnamefont {Yoshikawa}},\ }\bibfield  {title}
  {\enquote {\bibinfo {title} {Margin and energy dissipation of adiabatic
  quantum-flux-parametron logic at finite temperature},}\ }\href@noop {}
  {\bibfield  {journal} {\bibinfo  {journal} {IEEE transactions on applied
  superconductivity}\ }\textbf {\bibinfo {volume} {23}},\ \bibinfo {pages}
  {1700304--1700304} (\bibinfo {year} {2012})}\BibitemShut {NoStop}%
\bibitem [{\citenamefont {Takeuchi}, \citenamefont {Yamanashi},\ and\
  \citenamefont {Yoshikawa}(2017)}]{takeuchi2017reversibility}%
  \BibitemOpen
  \bibfield  {author} {\bibinfo {author} {\bibfnamefont {N.}~\bibnamefont
  {Takeuchi}}, \bibinfo {author} {\bibfnamefont {Y.}~\bibnamefont
  {Yamanashi}},\ and\ \bibinfo {author} {\bibfnamefont {N.}~\bibnamefont
  {Yoshikawa}},\ }\bibfield  {title} {\enquote {\bibinfo {title} {Reversibility
  and energy dissipation in adiabatic superconductor logic},}\ }\href@noop {}
  {\bibfield  {journal} {\bibinfo  {journal} {Scientific reports}\ }\textbf
  {\bibinfo {volume} {7}},\ \bibinfo {pages} {75} (\bibinfo {year}
  {2017})}\BibitemShut {NoStop}%
\bibitem [{\citenamefont {Yamae}, \citenamefont {Takeuchi},\ and\ \citenamefont
  {Yoshikawa}(2024)}]{yamae2024minimum}%
  \BibitemOpen
  \bibfield  {author} {\bibinfo {author} {\bibfnamefont {T.}~\bibnamefont
  {Yamae}}, \bibinfo {author} {\bibfnamefont {N.}~\bibnamefont {Takeuchi}},\
  and\ \bibinfo {author} {\bibfnamefont {N.}~\bibnamefont {Yoshikawa}},\
  }\bibfield  {title} {\enquote {\bibinfo {title} {Minimum energy dissipation
  required for information processing using adiabatic quantum-flux-parametron
  circuits},}\ }\href@noop {} {\bibfield  {journal} {\bibinfo  {journal}
  {Journal of Applied Physics}\ }\textbf {\bibinfo {volume} {135}} (\bibinfo
  {year} {2024})}\BibitemShut {NoStop}%
\bibitem [{\citenamefont {Ayala}\ \emph {et~al.}(2020)\citenamefont {Ayala},
  \citenamefont {Tanaka}, \citenamefont {Saito}, \citenamefont {Nozoe},
  \citenamefont {Takeuchi},\ and\ \citenamefont {Yoshikawa}}]{ayala2020mana}%
  \BibitemOpen
  \bibfield  {author} {\bibinfo {author} {\bibfnamefont {C.~L.}\ \bibnamefont
  {Ayala}}, \bibinfo {author} {\bibfnamefont {T.}~\bibnamefont {Tanaka}},
  \bibinfo {author} {\bibfnamefont {R.}~\bibnamefont {Saito}}, \bibinfo
  {author} {\bibfnamefont {M.}~\bibnamefont {Nozoe}}, \bibinfo {author}
  {\bibfnamefont {N.}~\bibnamefont {Takeuchi}},\ and\ \bibinfo {author}
  {\bibfnamefont {N.}~\bibnamefont {Yoshikawa}},\ }\bibfield  {title} {\enquote
  {\bibinfo {title} {Mana: A monolithic adiabatic integration architecture
  microprocessor using 1.4-zj/op unshunted superconductor {J}osephson junction
  devices},}\ }\href@noop {} {\bibfield  {journal} {\bibinfo  {journal} {IEEE
  Journal of Solid-State Circuits}\ }\textbf {\bibinfo {volume} {56}},\
  \bibinfo {pages} {1152--1165} (\bibinfo {year} {2020})}\BibitemShut {NoStop}%
\bibitem [{\citenamefont {Ko}\ and\ \citenamefont {Lee}(1992)}]{ko1992noise}%
  \BibitemOpen
  \bibfield  {author} {\bibinfo {author} {\bibfnamefont {H.~L.}\ \bibnamefont
  {Ko}}\ and\ \bibinfo {author} {\bibfnamefont {G.~S.}\ \bibnamefont {Lee}},\
  }\bibfield  {title} {\enquote {\bibinfo {title} {Noise analysis of the
  quantum flux parametron},}\ }\href@noop {} {\bibfield  {journal} {\bibinfo
  {journal} {IEEE transactions on applied superconductivity}\ }\textbf
  {\bibinfo {volume} {2}},\ \bibinfo {pages} {156--164} (\bibinfo {year}
  {1992})}\BibitemShut {NoStop}%
\bibitem [{\citenamefont {Fredkin}\ and\ \citenamefont
  {Toffoli}(1982)}]{fredkin1982conservative}%
  \BibitemOpen
  \bibfield  {author} {\bibinfo {author} {\bibfnamefont {E.}~\bibnamefont
  {Fredkin}}\ and\ \bibinfo {author} {\bibfnamefont {T.}~\bibnamefont
  {Toffoli}},\ }\bibfield  {title} {\enquote {\bibinfo {title} {Conservative
  logic},}\ }\href@noop {} {\bibfield  {journal} {\bibinfo  {journal}
  {International Journal of theoretical physics}\ }\textbf {\bibinfo {volume}
  {21}},\ \bibinfo {pages} {219--253} (\bibinfo {year} {1982})}\BibitemShut
  {NoStop}%
\bibitem [{\citenamefont {B{\'e}rut}\ \emph {et~al.}(2012)\citenamefont
  {B{\'e}rut}, \citenamefont {Arakelyan}, \citenamefont {Petrosyan},
  \citenamefont {Ciliberto}, \citenamefont {Dillenschneider},\ and\
  \citenamefont {Lutz}}]{berut2012experimental}%
  \BibitemOpen
  \bibfield  {author} {\bibinfo {author} {\bibfnamefont {A.}~\bibnamefont
  {B{\'e}rut}}, \bibinfo {author} {\bibfnamefont {A.}~\bibnamefont
  {Arakelyan}}, \bibinfo {author} {\bibfnamefont {A.}~\bibnamefont
  {Petrosyan}}, \bibinfo {author} {\bibfnamefont {S.}~\bibnamefont
  {Ciliberto}}, \bibinfo {author} {\bibfnamefont {R.}~\bibnamefont
  {Dillenschneider}},\ and\ \bibinfo {author} {\bibfnamefont {E.}~\bibnamefont
  {Lutz}},\ }\bibfield  {title} {\enquote {\bibinfo {title} {Experimental
  verification of {L}andauer’s principle linking information and
  thermodynamics},}\ }\href@noop {} {\bibfield  {journal} {\bibinfo  {journal}
  {Nature}\ }\textbf {\bibinfo {volume} {483}},\ \bibinfo {pages} {187--189}
  (\bibinfo {year} {2012})}\BibitemShut {NoStop}%
\bibitem [{\citenamefont {Bennett}(2003)}]{bennett2003notes}%
  \BibitemOpen
  \bibfield  {author} {\bibinfo {author} {\bibfnamefont {C.~H.}\ \bibnamefont
  {Bennett}},\ }\bibfield  {title} {\enquote {\bibinfo {title} {Notes on
  landauer's principle, reversible computation, and maxwell's demon},}\
  }\href@noop {} {\bibfield  {journal} {\bibinfo  {journal} {Studies In History
  and Philosophy of Science Part B: Studies In History and Philosophy of Modern
  Physics}\ }\textbf {\bibinfo {volume} {34}},\ \bibinfo {pages} {501--510}
  (\bibinfo {year} {2003})}\BibitemShut {NoStop}%
\bibitem [{\citenamefont {Herr}, \citenamefont {Josephsen},\ and\ \citenamefont
  {Herr}(2023)}]{herr2023superconducting}%
  \BibitemOpen
  \bibfield  {author} {\bibinfo {author} {\bibfnamefont {Q.}~\bibnamefont
  {Herr}}, \bibinfo {author} {\bibfnamefont {T.}~\bibnamefont {Josephsen}},\
  and\ \bibinfo {author} {\bibfnamefont {A.}~\bibnamefont {Herr}},\ }\bibfield
  {title} {\enquote {\bibinfo {title} {Superconducting pulse conserving logic
  and {J}osephson-{SRAM}},}\ }\href@noop {} {\bibfield  {journal} {\bibinfo
  {journal} {Applied Physics Letters}\ }\textbf {\bibinfo {volume} {122}}
  (\bibinfo {year} {2023})}\BibitemShut {NoStop}%
\bibitem [{\citenamefont {Berkley}\ \emph {et~al.}(2010)\citenamefont
  {Berkley}, \citenamefont {Johnson}, \citenamefont {Bunyk}, \citenamefont
  {Harris}, \citenamefont {Johansson}, \citenamefont {Lanting}, \citenamefont
  {Ladizinsky}, \citenamefont {Tolkacheva}, \citenamefont {Amin},\ and\
  \citenamefont {Rose}}]{berkley2010scalable}%
  \BibitemOpen
  \bibfield  {author} {\bibinfo {author} {\bibfnamefont {A.}~\bibnamefont
  {Berkley}}, \bibinfo {author} {\bibfnamefont {M.}~\bibnamefont {Johnson}},
  \bibinfo {author} {\bibfnamefont {P.}~\bibnamefont {Bunyk}}, \bibinfo
  {author} {\bibfnamefont {R.}~\bibnamefont {Harris}}, \bibinfo {author}
  {\bibfnamefont {J.}~\bibnamefont {Johansson}}, \bibinfo {author}
  {\bibfnamefont {T.}~\bibnamefont {Lanting}}, \bibinfo {author} {\bibfnamefont
  {E.}~\bibnamefont {Ladizinsky}}, \bibinfo {author} {\bibfnamefont
  {E.}~\bibnamefont {Tolkacheva}}, \bibinfo {author} {\bibfnamefont
  {M.}~\bibnamefont {Amin}},\ and\ \bibinfo {author} {\bibfnamefont
  {G.}~\bibnamefont {Rose}},\ }\bibfield  {title} {\enquote {\bibinfo {title}
  {A scalable readout system for a superconducting adiabatic quantum
  optimization system},}\ }\href@noop {} {\bibfield  {journal} {\bibinfo
  {journal} {Superconductor Science and Technology}\ }\textbf {\bibinfo
  {volume} {23}},\ \bibinfo {pages} {105014} (\bibinfo {year}
  {2010})}\BibitemShut {NoStop}%
\end{thebibliography}%

\end{document}